\documentclass[%
 reprint,longbibliography,preprintnumbers,
nofootinbib,
 amsmath,amssymb,
 aps
]{revtex4-1}
\pdfoutput=1
\usepackage{graphicx}
\usepackage[utf8]{inputenc}
\usepackage{flushend}
\usepackage{dcolumn}
\usepackage{bm}
\usepackage{balance}

\usepackage[normalem]{ulem}

\usepackage[colorlinks = true,
            linkcolor = blue,
            urlcolor  = blue,
            citecolor =green,
            anchorcolor = blue]{hyperref}
\usepackage{verbatim}
\usepackage{color,ulem}
\usepackage[english]{babel}

\usepackage[utf8]{inputenc}
\input Starburst.fd
\newcommand*\initfamily{\usefont{U}{Starburst}{xl}{n}}\initfamily

\newcommand{\beq}{\begin{eqnarray}}
\newcommand{\eeq}{\end{eqnarray}}
\usepackage{amsmath}
\usepackage{tikz}
\usetikzlibrary{decorations.pathmorphing}
\usetikzlibrary{shapes.misc}
\tikzset{cross/.style={cross out, draw=black, minimum size=8*(#1-\pgflinewidth), inner sep=0pt, outer sep=0pt},
cross/.default={1pt}}
\usetikzlibrary{patterns,math}
\begin{document}

\title{Phonon-confinement theory of thermal conductivity in ultrathin silicon films}

\author{\textbf{Alessio Zaccone}$^{1,2}$}%
 \email{alessio.zaccone@unimi.it}
 
 \vspace{1cm}

\affiliation{$^{1}$Department of Physics ``A. Pontremoli'', University of Milan, via Celoria 16,
20133 Milan, Italy.}
\affiliation{$^2$ 
Institut für Theoretische Physik, University of G{\"o}ttingen,
Friedrich-Hund-Platz 1,
37077 G{\"o}ttingen, Germany}

\begin{abstract}
The thermal properties of solids under nanoscale confinement are currently not understood at the atomic level. Recent numerical studies have highlighted the presence of a minimum in the thermal conductivity as a function of thickness for ultrathin films at a thickness about 1-2 nm, which cannot be described by existing theories. 
We develop a theoretical description of thin films which predicts a new physical law for heat transfer at the nanoscale. In particular, due to the strong redistribution of phonon momentum states in reciprocal space (with a transition from a spherical Debye surface to a different homotopy group $\mathbb{Z}$ at strong confinement), the low-energy phonon density of states no longer follows Debye's law but rather a cubic law with frequency, which then crosses over to Debye's law at a crossover frequency proportional to the average speed of sound of the material and inversely proportional to the film thickness. Concomitantly, this implies that the phonon population becomes dominated by low-energy phonons as confinement increases, which then leads to a higher thermal conductivity under extreme confinement. The theory is able to reproduce the thermal conductivity minimum in recent molecular simulations data for ultrathin silicon and provides useful guidelines so as to tune the minimum position based on the mechanical properties of the material.
\end{abstract}

\maketitle
\section{Introduction}
Understanding the thermodynamic and transport properties of nanoscale materials is important for both our fundamental understanding of condensed matter and also for many technological applications. Ultrathin films, in particular, are now routinely being fabricated with thickness of the order of few nanometers, such that confinement along the thin direction has a strong impact on the electronic, magnetic and thermal properties. 

Understanding heat transfer at the nanoscale has profound implications for electronics \cite{Liu2012}:
as transistors shrink, heat dissipation becomes a major bottleneck in high-performance computing. Furthermore, nanoscale heat transfer helps design materials and structures (like graphene or carbon nanotubes) to manage heat more efficiently \cite{Balandin2011,Balandin2020}. In quantum computing systems, precise thermal control at the nanoscale is essential to maintain coherence in quantum bits.
In superconducting materials, such as in Superconducting Nanowire
Single-Photon Detectors (SNSPDs) that find application in
space communications, the phonon-mediated heat transfer is a key parameter to control the detection quantum yield \cite{nanowire,Sidorova_2023}.
In energy harvesting applications, nanoscale materials can dramatically enhance thermoelectric performance by reducing thermal conductivity while maintaining high electrical conductivity, which improves the thermoelectric figure of merit \cite{Tian2014}.
Finally, understanding and controlling the thermal conductivity of ultrathin films is key to engineer layered and van der Waals materials where the heat transfer is quasi-2D in the atomic layers.

Theoretical frameworks based on the Boltzmann equation, including scattering of phonons that bump off the boundaries of the film, as described by the Fuchs-Sondheimer (FS) theory \cite{Fuchs_1938,Sond_1952}, remain invaluable to explain the increase of thermal conductivity with increasing film thickness in the range of tens to hundreds of nanometers \cite{Amon}. However, recent numerical investigations have revealed new unexpected behaviors upon further shrinking the thickness, that cannot be described by traditional theories. In particular, molecular simulations have highlighted the presence of a minimum in the thermal conductivity of ultrathin silicon films at about 1-2 nm thickness \cite{McGaughey}. A similar behaviour, although without a clear minimum and much shallower, has been reported for argon films \cite{McGaughey}.

As of today, the existing theoretical frameworks cannot explain these observation of the minimum in thermal conductivity at 1-2 nm thickness in silicon films and the flattening behaviour in argon films. Here we apply the nanoconfinement description of phonons in ultrathin few-layer films based on recent developments \cite{Yu_2022,Zaccone_heat} and derive a new fundamental framework for the thermal conductivity of ultrathin films that fully takes into account the restriction in the reciprocal space occupancy of phonons due to confinement \cite{Balandin1998}. The new theory is verified in qualitative comparison with recent molecular dynamics (MD) simulations data in terms of the predicted thickness dependence.

In particular, the origin of the minimum is traced back to a crossover in the phonon density of states from a confinement-induced cubic law in frequency to the quadratic Debye law. This effect, in turn, originates from a redistribution of phonon momentum states in reciprocal space associated with a transition from the trivial Debye-sphere homotopy group to a non-trivial manifold which belongs to the homotopy class $\mathbb{Z}$.

\section{Model assumptions}
We should emphasize the physical assumptions of the theory to be presented below. First of all, we shall always treat the ultra-thin films as 3D nanomaterials, where the atomic vibrations occur in all three spatial directions. This is true also for ultra-thin films of 1-2 nm thickness which possess a few atomic layers also along the confinement direction. Hence, we shall not consider exactly or ''ideal" 2D systems which are monolayers where the atoms vibrate only in the plane, since their thickness is much lower than the thickness considered in this paper.
Indeed, for a Si monolayer, the thickness ranges from just 0.4-0.5 {\AA} in silicene to about 3-4 {\AA} for Si(111) monolayers \cite{Lee2013,Zhao2016,Pawlak2020,Chen2015}. In what follows, instead, we will focus on films with thickness larger than $\sim 1$ nm, which thus contain at least 2-3 atomic layers. These systems are of course, fully 3D, from the point of view of phonon dynamics, as atomic vibrations take place in all three spatial directions. 

Furthermore, while the theory reported here is agnostic in terms of the type of vibrational excitations that contribute to the density of states, we argue that flexural modes do not change the basic form of the DOS ($\sim \omega^3$) dictated by nanoconfinement, as observed experimentally and in simulations for ultra-thin films \cite{Yu_2022,McGaughey,Guo_2023}. In these systems, the characteristic square-root of frequency trend of the low-energy phonon density of states (DOS) induced by flexural modes \cite{Kosevich} is never observed.

\section{Derivation}
The phonon thermal conductivity of a solid can be evaluated according to the Debye-Peierls or Debye-Callaway theory as:
\begin{equation}
    \kappa = \frac{1}{3}\int_0^{\omega_D}C_v(\omega)v(\omega)^{2}\tau(\omega)d\omega \label{cond}
\end{equation}
where $C_v(\omega)$ is the frequency-dependent volumetric heat capacity, $v(\omega)$ is the phonon group velocity, and
$\tau(\omega)$ is the phonon scattering time. The latter receives contribution from various processes, that include: Rayleigh-type scattering due to isolated defects, anharmonic phonon-phonon scattering, boundary scattering etc. As we are going to specialize on a simple monoatomic system such as silicon, we can use the original Debye-Peierls model without accounting for optical phonons explicitly. Then, the spectral heat capacity is given by the energy of a mode with frequency $\omega$: $U = g(\omega)\langle n(\omega)\rangle \hbar \omega$, where $g(\omega)$ is the phonon density of states (DOS), and $\langle n(\omega)\rangle=\frac{\hbar \omega}{e^{\hbar \omega/k_{B}T}-1}$ is the Bose-Einstein occupation factor. It should be noted that, using the Bose-Einstein quantum statistics, does not imply any loss of generality because at room temperature the Bose-Einstein statistics reduces to the Boltzmann statistics. 
Taking the derivative of the internal energy of mode $\omega$ with respect to temperature $T$, and using the Debye DOS \cite{Kittel}: $g(\omega)=\frac{3V}{2\pi^2}\frac{\omega^2}{v^3}$, the spectral heat capacity per unit volume is then:
\begin{equation}
    C_v(\omega)=\frac{3\hbar^{2}}{2\pi^{2}v^{3}k_{B}T^{2}}\frac{\omega^{4}e^{\hbar \omega/k_{B}T}}{\left(e^{\hbar \omega/k_{B}T}-1\right)^{2}}
\end{equation}
where $v$ denotes the average sound velocity.

The average group velocity for acoustic phonons can be expressed (using the Born-von Karman periodic boundary condition) as: $v(\omega)=v \sqrt{1-(\omega/\omega_D)^2}$, where $\omega_D$ is the Debye frequency of the material (i.e. the maximum phonon frequency in the system). 

Finally, the frequency-dependent inverse of the mean free path of the phonons, which is proportional to the inverse of the average relaxation time, is given by \cite{Broido,Walker_Pohl,Braun,Callaway}:
\begin{equation}
\Lambda^{-1}_{bulk}(\omega) \propto \tau^{-1}(\omega)=\frac{v}{L}+A \omega^4 + B \omega^2 e^{-C/T} \label{path}
\end{equation}
where the first term on the r.h.s. is a boundary scattering term \cite{Amon,Larkin,Braun} which suffices to describe size-effects in thin films with thickness $L > 10$ nm. The second term on the r.h.s. represents the Rayleigh-type scattering of phonons by diluted defects and impurities (a similar quartic form is found also for more generic atomic-scale disorder \cite{Baggioli_damping}).
The last term on the r.h.s. is the anharmonic contribution from inelastic three-phonon (Umklapp) scattering, famously theorized by Peierls \cite{Peierls1955} and subsequently computed by Klemens \cite{Klemens}.

Typical values of the scattering parameters for crystalline silicon based on several data sets in the literature \cite{Braun} are: $A= 1.82 \times 10^{-45}$ s$^{3}$, $B=2.8 \times  10^{-19}$ s K$^{-1}$, $C=182$ K. 

The standard theory outlined above provides a reasonable description of thermal conductivity of silicon, as well as other insulators and semiconductors \cite{Toberer,Ren}. Furthermore, the boundary-scattering term mentioned above, can be  supplemented by the Fuchs-Sondheimer (FS) interface scattering theory \cite{Fuchs_1938,Sond_1952} to include a detailed description of phonons bumping into the edges of the film, leading to more energy dissipation and, hence, to lower thermal conductivity \cite{Amon}.

\begin{figure}[h]
\centering
\includegraphics[width=\linewidth]{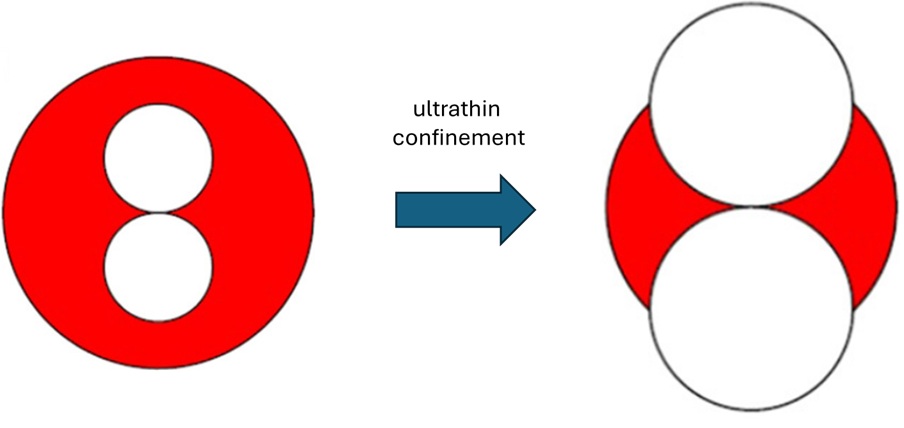}
\caption{Left: diametral section of the corresponding geometry of $k$-space, where the Debye sphere (of radius $k_{D}$, in red) contains two symmetric "hollow" spheres (hole pockets of radius $\pi/L$, in white) of forbidden states. These are phonon states in $k$-space that remain unoccupied due to confinement along the $z$-axis of the thin film in real space. Upon further decreasing the film thickness $L$, eventually the Debye spherical surface gets distorted, as shown in the right figure. At this point the phonon momentum states in reciprocal space are strongly redistributed, with a spreading out of states on the distorted Debye surface and many states being pushed towards the core of the red figure, resulting in a faster growth of the number of states with $k$, hence with $\omega$, explaining the higher, cubic, exponent, of the DOS derived in \cite{Yu_2022}. Reproduced from [R. Travaglino and A. Zaccone, Journal of Applied Physics 133, 033901
(2023)], with the permission of AIP Publishing. }
\label{fig1}
\end{figure}

However, additional effects arise when the film thickness $L$ drops below 10 nm, which are not taken into account by the above theory framework.
These new effects are brought about by the spatial confinement of the quasiparticles, and can affect both thermal \cite{Zaccone_heat} and electronic transport \cite{Zaccone_cond}, as well as the superconductivity \cite{Travaglino_2023,Ummarino}.
In a nutshell, phonons described as quantum waves have a certain wavelength. Along the confined ($z$) direction of the film, only wavelengths smaller than the thickness $L$ are allowed to exist and to populate the Debye sphere in reciprocal space. On the other hand, in the unconfined $x-y$ plane, phonons with any wavelength are allowed.
Along a generic direction identified by the angle $\theta$ (measured with respect to the confinement axis $z$), only phonons with wavelength:
\begin{equation}
    \lambda < \lambda_{max}=\frac{L}{\cos \theta}
\end{equation}
are allowed to populate the Debye sphere in $k$-space \cite{Zaccone_rev}. Phonons with larger $\lambda$ are forbidden and the corresponding points in $k$-space form two symmetric hole-pocket spheres of prohibited states within the Debye sphere, cf. Fig.\ref{fig1} and Figs. 1-2 in \cite{Phillips}.
As long as these two hole-pocket spheres are within the surface of the Debye sphere, the DOS is still provided by the Debye DOS quoted above. However, as the thickness $L$ is decreased, the two hole-pocket "hollow" spheres grow up to the point that they come out of the Debye sphere (right figure in Fig. \ref{fig1}). At this point, the DOS is no longer given by the Debye DOS but can still be derived in simple analytical form, as follows. 

The volume occupied in $k$-space (red region in the sphere diametral section on the right of Fig. \ref{fig1}) is determined by the volume of the Debye sphere minus the overlapping region shared with the two spheres of forbidden states. This overlapping volume, denoted as $V_{inter}$, corresponds to the intersection between the $L$-dependent volume of the two white spheres shown in Fig. \ref{fig1} and the red Debye sphere. It can be computed exactly using standard methods from solid geometry \cite{Travaglino_2022}. Specifically, the calculation involves integrating the cross-sectional areas of stacked disks along the $k_z$-axis (generally, the volume of a cylindrically symmetric body is obtained by summing the areas of infinitesimally thin disks stacked along one axis; because there are infinitely many such disks, this summation becomes an integral) \cite{Travaglino_2022,Travaglino_2023}. The precise result yields $V_{inter} = \frac{4\pi k^3}{3} - \frac{L k^4}{2}$, and therefore, the occupied volume in $k$-space is given by \cite{Travaglino_2022,Travaglino_2023}:
\begin{equation}
Vol_{k} = \frac{4\pi k^3}{3} - V_{inter} = \frac{L k^4}{2}.
\end{equation}
The corresponding number of states in $k$-space with $k < k'$ is then:
\begin{equation}
N(k < k') = \frac{V}{(2\pi)^3} \frac{L k^4}{2}.
\end{equation}
From this, the phonon density of states in the thin film can be directly obtained:
\begin{equation}
g(\omega) = \frac{d}{d\omega} N(\omega < \omega') = \frac{V}{4\pi^3} L \frac{\omega^3}{v^4} \label{VDOS}
\end{equation}
which shows a cubic dependence on frequency, $\sim \omega^3$, consistent with both experimental results (from inelastic neutron scattering) and molecular dynamics simulations reported in Ref. \cite{Yu_2022}. Notably, this $\omega^3$ behavior is observed in both crystalline and fully amorphous thin films.
Furthermore, this $\omega^3$ law is approximately observed in MD simulations of ultrathin silicon films and ultrathin argon films \cite{Turney,McGaughey}. This cubic DOS for ultrathin films leads to a characteristic $\sim T^4$ of the heat capacity with temperature at low temperatures (instead of Debye's $\sim T^3$), which appears in qualitative agreement with the available data \cite{Zaccone_heat}.

The above DOS is for just one phonon polarization, and a factor of three has to be implemented when computing the total internal energy $U$ and the heat capacity \cite{Kittel}.

As already derived and demonstrated in Ref. \cite{Yu_2022}, the above DOS crosses over into the Debye DOS at the crossover frequency value: $\omega_{\times}=\frac{2\pi}{L}v$. This crossover was confirmed experimentally and numerically in \cite{Yu_2022}.
For a thin film which is several nanometers thick, the crossover frequency $\omega_{\times}$ is on the order of few TeraHertz (THz), hence about an order of magnitude lower than the Debye frequency  $\omega_D$.
Hence, the DOS for ultrathin films can be schematically described as follows \cite{Yu_2022}:
\begin{eqnarray}
  g (\omega) = \left\{\begin{array}{ll}
    \frac{3V}{4\pi^{3}} L  \frac{\omega^{3}}{v^4}, & \omega < \omega_{\times}\\
   \frac{3V}{2\pi^2}\frac{\omega^2}{v^3}, & \omega > \omega_{\times}.
  \end{array}\right.  \label{cubic}
\end{eqnarray}
This form of the DOS, besides having been extensively verified against atomistic simulations and experiments in \cite{Yu_2022}, also reproduces the observations of Refs. \cite{Turney,McGaughey} on MD simulations of ultrathin silicon crystals and of argon crystals. In those works it was clearly seen, in the DOS obtained in the simulations, that the DOS at low frequency displays a higher density of modes with a higher exponent than Debye's quadratic law, compatible with the cubic law above resulting from phonon confinement. At the crossover frequency $\omega_{\times}=\frac{2\pi}{L}v$, the cubic law melds into the quadratic law briefly before the onset of van Hove peaks and other typical higher-frequency features of the DOS which are dominated by optical phonons, (not considered here explicitly) \cite{Turney,McGaughey}.

From a physical point of view, the cubic law for the DOS at low frequency in Eq. \eqref{cubic} resulting from phonon confinement, reflects the redistribution of momentum states of the phonons in $k$-space, as the confinement is increased, as schematically depicted in Fig. \ref{fig1}. As the two hollow spheres of forbidden states grow out of the Debye sphere, many phonon states are pushed towards the inner core of the occupied volume (in red, in the figure) thus leading to a faster growing number of states with $k$, and hence to a faster growing number of states with frequency $\omega$, indeed with a higher, cubic, exponent of the DOS. 

Correspondingly, for the spectral heat capacity, we obtain:
\begin{eqnarray}
  C_v(\omega) = \left\{\begin{array}{ll}
    \frac{3\hbar^{2} L}{4\pi^{3}v^4 k_{B}T^{2}}   \frac{\omega^{5}e^{\hbar \omega/k_{B}T}}{\left(e^{\hbar \omega/k_{B}T}-1\right)^{2}}, & \omega < \omega_{\times}\\
   \frac{3\hbar^{2}}{2\pi^{2}v^{3}k_{B}T^{2}}\frac{\omega^{4}e^{\hbar \omega/k_{B}T}}{\left(e^{\hbar \omega/k_{B}T}-1\right)^{2}}, & \omega > \omega_{\times}.
  \end{array}\right.  \label{spectral}
\end{eqnarray}
with $\omega_{\times}=2 \pi v/L$.

This physical model thus explains how the confinement leads to a \emph{relative} increase in the density of low-frequency vibrational modes, as indeed observed in simulations and experiments \cite{Yu_2022,Turney,McGaughey}. In particular, the excess of vibrational modes over the Debye prediction is controlled by the crossover in Eq. \eqref{cubic} occurring at $\omega_{\times}=\frac{2\pi}{L}v$, itself a function of the film thickness $L$. Clearly, as $L$ decreases, the extension of the cubic-law regime in the DOS grows, and, with it, the relative density of low-frequency phonons (compared to high-frequency ones). As is well known \cite{Ziman}, low-frequency phonons dominate the thermal conductivity, because they have a larger mean free path and are more resilient to scattering. The fact that the overall phonon population distribution is shifted towards low-frequency phonons, thus leads to an increase of thermal conductivity with increasing confinement (decreasing thickness $L$) in this regime. 
We should also remark that these low-frequency phonons whose importance grows as $L$ decreases are all, practically, in-plane phonons since long wavelengths can mostly be accommodated in the $x-y$ plane of the film. Hence, this effect is going to be important for the in-plane thermal conductivity.

We should also emphasize that it is the relative contribution of low-frequency phonons, not their absolute count, which matters for the thermal conductivity. Ultimately, this is because thermal conductivity is a weighted integral, not a particle count. The integral sums how effectively each frequency channel transports heat, and the effectiveness is determined by the transport weight $v(\omega)^2\tau(\omega)$, not by the absolute number of phonons at that frequency. Hence, the relative contribution of low-frequency phonons matters because thermal conductivity is determined not by how many phonons exist, but by how strongly each frequency channel is weighted by its transport efficiency as encoded in the product $v(\omega)^2\tau(\omega)$ in the integrand; low-frequency acoustic modes have overwhelmingly larger efficiency, so even a small fraction of them dominates the heat flow.

To quantitatively test the above theory, we also include the FS interface scattering, following the analytical formulation of Ref. \cite{Cuffe}. This is implemented as a correction function that multiplies the bulk mean free path $\Lambda_{bulk}$:
\begin{align}
S\left( \frac{\Lambda_{bulk}}{L} \right)& = 1 - \frac{3}{8} \frac{\Lambda_{bulk}}{L} \nonumber\\
&+ \frac{3}{2} \frac{\Lambda_{bulk}}{L} \int_1^{\infty} \left( \frac{1}{t^3} - \frac{1}{t^5} \right) 
e^{- \frac{L}{\Lambda_{bulk}} t} \, dt \label{FS}
\end{align}
such that the thickness-dependent mean free path according to FS theory is: $\Lambda = S\left( \frac{\Lambda_{bulk}}{L} \right)\Lambda_{bulk}$. Considering that $\tau = \Lambda/v$, we thus rewrite Eq. \eqref{cond} as \cite{Cuffe}:
\begin{equation}
    \kappa = \frac{1}{3}\int_0^{\omega_D}C_v(\omega)v(\omega)S\left( \frac{\Lambda_{bulk}}{L} \right)\Lambda_{bulk}(\omega)d\omega. \label{final}
\end{equation}

\section{Physical intuition behind the transition from phonon hole-pockets to a distorted Debye surface}

When an ultrathin film is confined along one spatial direction, the standing-wave condition 
$|k_z| \ge \pi/L$ suppresses all vibrational modes whose out-of-plane wavelength exceeds the 
film thickness $L$. In reciprocal space, this eliminates an entire set of states near $k_z=0$, 
corresponding geometrically to two spherical regions of radius $\pi/L$ that are carved out of 
the interior of the Debye sphere. For moderately confined films, these ``hole pockets'' remain 
entirely within the Debye sphere, and the overall structure of the allowed $k$-space volume is 
still close to a spherical shell with internal cavities, so that the density of states retains 
approximately the Debye form $g(\omega)\!\sim\!\omega^{2}$.

As the film is thinned further, the radius of the forbidden pockets grows until these cavities 
pierce the boundary of the Debye sphere. Beyond this point, the geometry of the allowed 
$k$-space region is no longer a sphere with removed interior volume: the Debye surface itself 
is reshaped, becoming strongly compressed along the confinement direction and expanded in the 
transverse directions. Phonon states that would normally populate a spherical shell are forced 
onto this distorted manifold, leading to a faster growth of the number of states with increasing 
$k$ and thereby producing a low-frequency density of states that scales as $g(\omega)\!\sim\!\omega^{3}$ 
instead of the Debye $\omega^{2}$ law.

This geometric transition in reciprocal space has direct consequences for heat transport. The 
redistribution of vibrational states toward low frequencies enhances the relative population of 
long-wavelength acoustic modes, which possess higher group velocities and longer mean free paths. 
Thus, the distortion of the Debye surface under strong confinement naturally leads to an increase 
in the contribution of low-frequency phonons to the thermal conductivity, providing the physical 
basis for the non-monotonic thickness dependence observed in ultrathin silicon films.

\section{Negligible contribution of optical phonons to heat transport}
In crystalline semiconductors such as silicon, the lattice thermal conductivity is overwhelmingly dominated by acoustic phonons. This can be understood by considering several factors. 

First, acoustic modes have a nearly linear dispersion near the Brillouin zone center and therefore possess much higher group velocities (on the order of $5000$--$9000~\text{m/s}$) compared to the relatively flat optical branches, whose group velocities are typically $\lesssim 10^2$--$10^3~\text{m/s}$. By recalling that the square of the group velocity appears inside the integral for the thermal conductivity, cf. Eq. \eqref{cond}, it is clear that acoustic phonons play the dominant role.

Second, while optical phonons lie at higher energies (e.g.\ $\sim 60~\text{meV}$ in Si) and are only partially populated at room temperature, acoustic phonons are numerous and thermally excited across the full spectrum. 

Third, optical phonons experience short lifetimes due to strong anharmonic decay, whereas acoustic phonons---particularly the long-wavelength modes---exhibit longer mean free paths and hence contribute far more effectively to heat transport. Quantitative branch-resolved Boltzmann transport calculations confirm this picture: at $300~\text{K}$, more than $95\%$ of the thermal conductivity of silicon arises from the three acoustic branches (LA $\sim 35\%$, TA$_1 \sim 28\%$, TA$_2 \sim 32\%$), while all optical branches together account for only about $5\%$ of the total \cite{Wang2014}.

For all these reasons, it is appropriate to focus the theoretical description on the role of acoustic phonons, since that is also the range where confinement effects are more important \cite{McGaughey}, and use an averaged Debye-type schematic description for the optical sector, as done above. 

\section{Additional phonon scattering mechanisms in ultrathin films}
Beyond the intrinsic phonon--phonon and diffuse boundary scattering processes
discussed above, a comprehensive description of thermal transport in
experimentally realized ultrathin films requires incorporating several
additional scattering channels that become increasingly important at the
nanoscale.  First, \emph{interface roughness scattering} represents a major
source of momentum relaxation in thin films grown on substrates or fabricated
via vapor-phase deposition.  Atomic-scale height fluctuations at free or buried
interfaces lead to wavelength-dependent partial specularity, with the specularity
parameter often modeled as $p(\lambda) \simeq \exp[-16\pi^3 \eta^2/\lambda^2]$,
where $\eta$ denotes the root-mean-square roughness \cite{Ziman, Soffer1967}.
Such roughness strongly suppresses cross-plane heat flow, as short-wavelength
phonons experience nearly fully diffuse scattering, while only long-wavelength
acoustic modes retain partial specularity.

Second, \emph{material defects} such as substitutional impurities, dopants,
vacancies, mass-disorder inhomogeneities, as well as structural disorder, introduce Rayleigh-type elastic
scattering, which produces a characteristic $\omega^4$ dependence of the
inverse relaxation time \cite{Klemens1955, Carruthers1961,Baggioli_2022}.  In doped silicon,
impurity scattering can dominate over anharmonic Umklapp processes across a wide
frequency range, substantially reducing the population of mid- and
high-frequency phonons.  Line defects such as dislocations also contribute
significantly: the long-range strain fields surrounding dislocations scatter
phonons via a relaxation time that depends sensitively on branch and frequency
\cite{Klemens1955b, Granato1966}.  Grain boundaries, common in polycrystalline
or nanocrystalline films, act as partially transmitting interfaces whose
reflection probability depends on phonon wavelength, incident angle, and
polarization; they therefore constitute an additional source of diffuse
scattering and mode conversion \cite{Mayadas1969}.

Finally, in supported films or multilayers, \emph{substrate and interfacial
coupling} provide further channels for phonon momentum relaxation.  Acoustic
mismatch, partial phonon transmission into the substrate, and the presence of
interfacial layers (e.g.\ native oxides) reduce the effective mean free path of
phonons impinging on the interface \cite{SwartzPohl1989}.  These mechanisms
alter the boundary conditions experienced by the confined modes and modify the
balance between in-plane and cross-plane heat transport.  Taken together, these
extrinsic scattering mechanisms---interface roughness, mass disorder, point and
line defects, grain boundaries, and coupling to the substrate---play an
essential role in determining the thermal conductivity of ultrathin films in
realistic experimental settings and must therefore be considered alongside the
intrinsic phonon-confinement effects described in this work. 

In the model presented here, effects due to disorder mentioned above are effectively taken into account by the Rayleigh-type $\omega^4$ contribution to the phonon scattering. The other effects discussed above can be easily integrated within the proposed model for future applications thereof to experimental data.

\section{Comparison with simulations data and discussion}
We now proceed to substitute the form of the spectral heat capacity, Eq. \eqref{spectral} that accounts for phonon confinement, together with Eq. \eqref{path} and \eqref{FS}, in Eq. \eqref{final}. Using the above quoted values of the parameters for $\Lambda_{bulk}$ valid for silicon, and its Debye frequency: $\omega_D = 13.8 \times 10^{13}$ rad/s, we obtain the theoretical prediction for the thermal conductivity of ultrathin silicon films reported in Fig. \ref{fig2} below.

\begin{figure}[h]
\centering
\includegraphics[width=\linewidth]{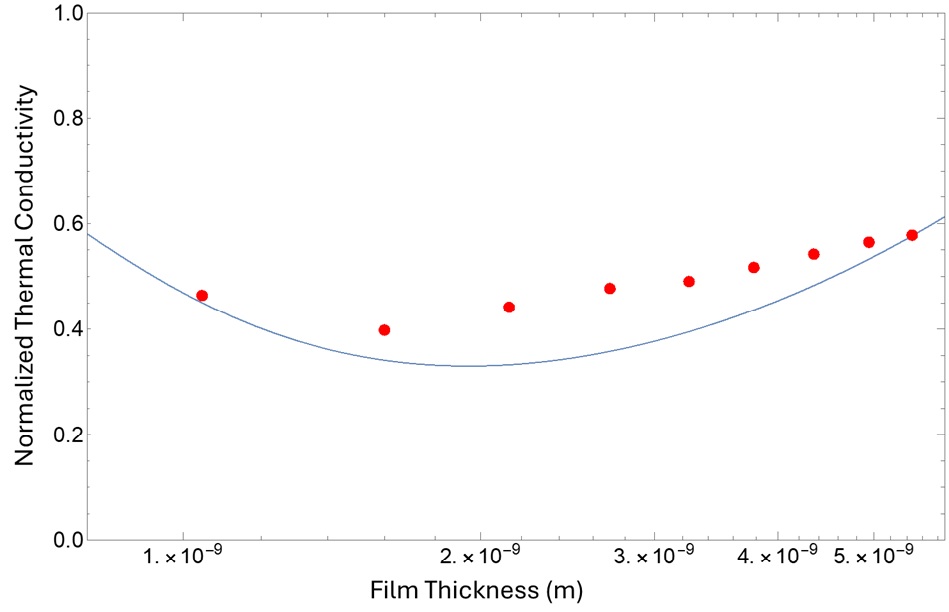}
\caption{Comparison between the prediction of the proposed theory (solid line) and MD simulations data (circles) for the thickness-dependent thermal conductivity of ultrathin silicon films at $T=200$ K. The MD data are from Ref. \cite{McGaughey}. Because the Debye-Peierls model used in the text is, of course, not fully quantitative, and the MD data from Ref. \cite{McGaughey} are also based on strong assumptions, the (thickness-independent) prefactors of $C_v(\omega)$ for $\omega < \omega_{\times}$ and for $\omega > \omega_{\times}$ in Eq. \eqref{spectral}, were adjusted, respectively, to $\frac{3\hbar^{2} }{4\pi^{3}v^4 k_{B}T^{2}}=9.5 \times 10^{-67}$ and $\frac{3\hbar^{2}}{2\pi^{2}v^{3}k_{B}T^{2}}=1.45 \times 10^{-63}$ (in SI units). These two values are not too far from the nominal values, at $T=200$ K, estimated to be $\frac{3\hbar^{2} }{4\pi^{3}v^4 k_{B}T^{2}}=2.8 \times 10^{-67}$ and $\frac{3\hbar^{2}}{2\pi^{2}v^{3}k_{B}T^{2}}=1.14 \times 10^{-62}$, with the average speed of sound for silicon, $v \approx 7000$ m/s. All other parameters are fixed and their literature values can be found in the text. The MD simulations data are normalized by the bulk value of the MD simulations, 414 W/m-K, while the theory line is normalized by an overall factor 9.2. }
\label{fig2}
\end{figure}

The crucial ingredient in the above model, allowing it to reproduce the minimum also observed in the simulated thermal conductivity of ultrathin silicon, is the cubic DOS at low frequency and its crossing over into the Debye DOS at $\omega_{\times}=2 \pi v/L$. Without this form of the DOS, we have checked that the traditional FS interface scattering mechanism is unable to produce the minimum.

The present theoretical model is the first, to our knowledge, which can predict a minimum in the thermal conductivy of ultrathin silicon as observed in MD simulations. It opens up the way for the rational control and tuning of thermal conductivity in ultrathin and layered materials. For example, the above model explains that the minimum depends crucially on the competition between the cubic regime of the DOS and the Debye quadratic regime, which is controlled by the crossover frequency $\omega_{\times}=2 \pi v/L$, a function of the average speed of sound $v$ of the material. The larger the extension of the cubic DOS in the vibrational spectrum, the larger the enhancement of the thermal conductivity in the ultrathin film, and the minimum gets more pronounced and shifted towards larger $L$ values.
So for example, for a mechanically softer material (with lower elastic moduli and hence lower $v$) such as the Lennard-Jones argon crystal, the above theory predicts that the minimum is shallower, compared to silicon which is a much harder material with much larger $v$. This theoretical prediction is fully consistent with the simulations data for argon in Ref. \cite{McGaughey}.

While different assumptions are used in the simulations of Ref. \cite{McGaughey} and in the present analytical theory, the presence of the minimum appears to be a robust feature, as well as the theoretical prediction of thermal conductivity increasing with film thickness for films thicker than 2 nm, as observed in many different materials \cite{Park2025}. 

\section{Conclusions}
In summary, we have derived an analytical theory for the thermal conductivity of ultrathin films based on a nanoconfinement model for vibrational modes. Due to the nanometric confinement along one spatial direction, there exist spherical hole pockets inside the Debye sphere (Fig. \ref{fig1}), which are associated with suppressed modes in the acoustic to THz regime of the vibrational spectrum. This, in turn, gives rise to a redistribution of modes on the distorted Debye surface, with a modified $\sim \omega^3$ density of states, that crosses over to the $\sim \omega^2$ Debye law at a crossover frequency $\omega_{\times}$ that is proportional to the average speed of sound and inversely proportional to the film thickness.
In other words, the relative importance of the low-energy phonon population over the more high-energy phonons, increases upon decreasing the film thickness into a strong confinement regime of a few nanometers. Below a point of minimum at about 2 nm, the low-energy phonon dominance is such that the thermal conductivity begins to increase with further confinement, because low-energy phonons are more effective contributors to the thermal conductivity. 

Instead, at larger thickness values than the point of minimum, the thermal conductivity increases monotonically with the thickness $L$ simply because more phonon modes are available to carry heat (as the two hole-pocket spheres in the left panel of Fig. \ref{fig1} keep shrinking with increasing $L$). This is consistent with the linear increase of the specific heat with $L$, $C_v(\omega) \sim L$, predicted by the phonon-confinement theory in \cite{Zaccone_heat}.

In future work, this theory can be extended in several directions such as the inclusion of phonon transmission factors between the thin film and the substrates, by systematic studying the effect of the various parameters in the perspective of developing effective phonon engineering strategies for nanomaterials. The present framework can also be applied to experimental data in the thicker range $5 < L < 500$ nm, as reported in \cite{Cuffe,ChavezAngel2014,VegaFlick2016}, in future work. Also, it will be interesting to apply the theory to more exotic materials such as few-layer 2D van der Waals materials,
graphene, BN, MoS$_2$, and amorphous and crystalline membranes, e.g. by including flexural and Lamb modes \cite{Kuhn} and disorder effects in the DOS \cite{PRR}. Important applications of the proposed framework to heat management of quantum computing systems are also envisaged.

\section*{Author Declarations}
The authors have no conflicts to disclose.

\section*{Data Availability}
Data sharing is not applicable to this article as no new data were created or analyzed in this study. 

\subsection*{Acknowledgments} 
A.Z. gratefully acknowledges funding from the European Union through Horizon Europe ERC Grant number: 101043968 ``Multimech'', and from US Army Research Office through contract nr. W911NF-22-2-0256. Several discussions with M. Sidorova and A. Semenov are gratefully acknowledged.
\bibliographystyle{apsrev4-1}

\bibliography{refs}

\end{document}